\documentclass{article}
 \usepackage{amsmath}
 \usepackage{authblk}
\usepackage[pdftex]{graphicx}
\usepackage{epstopdf}

\usepackage[usenames,dvipsnames,svgnames,table]{xcolor}
\newcommand{\half}{\ensuremath{^1\!\!/\!_2}}    %   a simple 1/2 symbol
\newcommand{\quarter}{\ensuremath{^1\!\!/\!_4}}    %   a simple 1/4 symbol

\begin{document}

\title{Atomistic-Object Kinetic Monte Carlo simulations of irradiation damage in tungsten}

    \author[1]{D.R. Mason\thanks{daniel.mason@ukaea.uk}}
    \author[2]{A.E. Sand}
    \author[1]{S.L. Dudarev}
    \affil[1]{CCFE, Culham Centre for Fusion Energy, Abingdon, Oxfordshire OX14~3DB, United Kingdom}
    \affil[2]{Department of Physics, University of Helsinki, P.O. Box 43, FI-00014, Helsinki, Finland}

\maketitle\thispagestyle{empty}

    \begin{abstract}
We describe the development of a new object kinetic Monte Carlo code where the elementary defect objects are off-lattice atomistic configurations.
Atomic-level transitions are used to transform and translate objects, to split objects and to merge them together.
This gradually constructs a database of atomic configurations- a set of relevant defect objects and their possible events generated on-the-fly.
Elastic interactions are handled within objects with empirical potentials at short distances, and between spatially distinct objects using the dipole tensor formalism.
The model is shown to evolve mobile interstitial clusters in tungsten faster than an equivalent molecular dynamics simulation, even at elevated temperatures.
We apply the model to the evolution of complex defects generated using molecular dynamics simulations of primary radiation damage in tungsten.
We show that we can evolve defect structures formed in cascade simulations to experimentally observable timescales of seconds while retaining atomistic detail. 
We conclude that the first few nanoseconds of simulation following cascade initiation would be better performed using molecular dynamics, as this will capture some of the near-temperature-independent evolution of small highly-mobile interstitial clusters.
We also conclude that, for the 20keV PKA cascades annealing simulations considered here, internal relaxations of sessile objects difficult to capture using conventional object KMC with idealised object geometries establish the conditions for long timescale evolution.
    \end{abstract}

    \section{Introduction}

    Nuclear materials for Gen IV fission and fusion power stations will be required to operate under unprecedented irradiation fluence.
The successful retention of good thermal and mechanical properties depends on the balance between damage generation due to neutron irradiation and restoration through defect recombination.
Predicting the evolution of defect clusters is therefore of utmost importance.

Two popular tools for characterizing defects, transmission electron microscopy (TEM)\cite{Jenkins2001,kirk_JMR2015}, atom-probe tomography (APT) \cite{Edmondson_MicAndMic2015} are very successful techniques for identifying (respectively) nanoscale dislocation loops and precipitates, but both struggle to identify the smallest point defects and clusters.
This is a significant omission, as state-of-the-art STEM microscopy\cite{Liu_Acta2017}, Field-Ion Microscopy \cite{Dagan_Ultramicroscopy2015}, and molecular dynamics (MD)\cite{Stoller_JNM1999,Setyawan_JNM2015} suggest that the majority of damage produced in primary damage cascades should be invisibly small defects.

But MD has also shown us that some defects generated during the heat spike phase of cascades are not small - a power-law size-frequency  distribution of defects has been found in simulations of irradiated tungsten and iron\cite{Sand_EPL2013,Sand_EPL2016}, and confirmed in experimental observations of self-ion irradiated tungsten\cite{Yi_EPL2015}.
Recently the spatial distribution\cite{Mason_Acta2018} and count per incident ion\cite{Mason_EPL2018} of larger defect clusters has also been measured in MD simulations and \emph{in situ} irradiations at cryogenic temperature, giving a characteristic length-scale of a single primary damage cascade for 150keV PKA ions in W of order one nanometre.
This proves that large defects are generated sufficiently close together that the interaction between them is a significant driving force on their subsequent evolution\cite{Mason_JPCM2014,Dudarev_PRB2010}.

To model the evolution of nanoscale defects to experimentally observable timescales at elevated temperature, we often turn to cluster dynamics (CD)\cite{Marian_JNM2011}, or object kinetic Monte Carlo (okMC)\cite{Domain_JNM2004,Becquart_JNM2009,MartinBragado_CPC2013}.
In the limit of no spatial correlation, ie defects produced homogeneously and with no elastic interactions, CD and okMC give essentially the same results\cite{Stoller_JNM2008}.
However, if defect objects are produced in a spatially correlated manner, then this could lead to some rapid recombination in okMC \cite{Jourdan_PRB2012,Nandipati_JNM2015}.
A challenge therefore is how to handle the spatial correlation in defect production- MD will give this information in the form of atomic positions, but okMC typically simplifies complex configurations to a small handful of parameters describing size, orientation and position of the defect.
It is not obvious that the complex atomic configurations seen in MD, particularly in overlapping cascades, should resolve quickly to prismatic loops and clusters\cite{San18,Granberg_EPL2017}.
Furthermore it may be the case that if this resolution to simpler forms does occur, it is because of the mutual proximity of defects.
okMC often relies on simple capture radii to determine whether clusters react, which may not correctly describe self-climb\cite{Swinburne_SciRep2016} or correlated atom motion in a dislocation core\cite{Dudarev_PM2003}.
Finally we note that including elastic interactions into okMC shows that dislocation loops can be trapped to experimentally observable times by drawing each other into favourable positions in their stress fields\cite{Mason_JPCM2014}.
The elastic interaction is most pronounced when loops are close together, which is exactly when the assumptions of okMC starts to break down.

On-the-fly kinetic Monte Carlo methods\cite{Beland_PRE2011,Xu_JPCM2012,Beland_CMS2015} are sometimes used to evolve arbitrarily complex defects, as they work by searching the configuration space of atomic positions for saddle points between atomic configurations\cite{Machado-Charry_JCP2011}.
Accelerated MD can also be used to seek transitions at high temperatures and then map back to transition rates at the desired low temperature\cite{Perez_ARCC2009,Swinburne_PRM2018}.
To find a barrier requires a considerable calculation effort, but when seeking infrequent events over thermally activated barriers, these methods are extremely powerful.
For very low barriers, of the order 0.1eV or less, such as might be seen for interstitial cluster movement, they struggle to compete with MD: put simply kMC methods are stochastic and so sample a range of local minima, whereas MD is deterministic and so moves from one minimum to the next.
Irradiation damage cascades give us very low barriers with quasi-independent interstitials, mid-range barriers for strongly interacting clusters and high barriers for vacancy clusters. All must be treated consistently.

In this paper we develop okMC with arbitrarily complex off-lattice atomic configurations.
The okMC framework constructs a database of previously visited objects and transitions, minimising recalculation.
However, rather than search for a comprehensive set of saddle points on-the-fly\cite{Henkelman2002}, we use a prescribed set of correlated atomic moves based on those well-known in MD and coarse-grained simulations, and well-parameterized by DFT.
We discuss the advantages and disadvantages of this simplified scheme.
We also demonstrate how to incorporate elastic interactions efficiently in an off-lattice kMC code.

In section \ref{individualDefects} we apply the model to some simple test cases to demonstrate correct and efficient running, and in section \ref{MDcascades} we apply the model directly to radiation damage cascades generated using MD.
We conclude that very mobile interstitial clusters may be rapidly lost to the boundaries of the simulation cell, or may collide.
If they do collide, then a complex sessile interstitial cluster may be formed first, which then later relaxes to a low-energy mobile cluster.
These effects would not be seen in earlier okMC models which use simple rules to determine the result of collisions, for instance taking only the experimental \textit{in situ} observation that when loops collide, the resultant loop takes the Burgers vector of the larger\cite{Arakawa_Acta2011}. 
%This is consistent both with the proposed mechanism of Marian et al \cite{Marian_PRL2002,Xu_PRL2013} that two $\half \langle 111 \rangle$ loops can collide to form a single $\langle 100 \rangle$ loop, and the experimental \textit{in situ} observation that when loops collide, the resultant loop takes the Burgers vector of the larger\cite{Arakawa_Acta2011} - our simulations are exploring both the short collision timescale and the longer relaxation timescale and only the latter is seen in experiment.
Our results using okMC are, however, in accord with previous MD observations of sessile defects formed by the collision of glissile interstitial defects \cite{Terentyev_PRL2008,Puigvi_PM2007,Anento_JNM2008}.
By capturing glissile to sessile to glissile transitions within an okMC framework, we are able to observe long-timescale detrapping mechanisms at microseconds and beyond, but which can preserve a population of interstitial defects in the microstructure.

    \section{Atomistic-Object Kinetic Monte Carlo}

            \label{okMC}

            The code we describe here could be described either as atomistic or object kinetic Monte Carlo, and either as having transitions computed on-the-fly or predetermined.
            As these terms are common in the kMC literature, and can appear to be mutually exclusive, we will first clarify our meaning.

            In materials science, particularly when discussing nano-scale defects in metals, it is common to talk about point defects, clusters, dislocation loops, voids etc.
            In doing so we are implicitly stating that such defects have a spatial localisation ( they exist as quasi-independent objects within a crystal ), and a temporal persistence ( they are metastable, rather than ephemeral atomic configurations ).
            This is the theoretical basis of object Kinetic Monte Carlo and cluster dynamics - we define the nano-scale defects as objects, and rules determine their dynamic evolution.
            An okMC state can be completely specified by the types of object, and their positions.
            An okMC model consists of the current state, together with rules for evolving the objects.
            A clear exposition of okMC is given in \cite{MartinBragado_CPC2013}.

            Often defect objects will be defined with a very small number of parameters - for example a void may be given a position and number of vacancies contained.
            The small number of parameters used is for convenience only, there is no \emph{a priori} reason why a void should not be given additional parameters defining its shape, if these could be meaningfully employed in the dynamics.
            The logical limit of object Kinetic Monte Carlo is to describe each object with atomic resolution: provided each atomic configuration has a padding boundary of perfect crystal, the defect regions are spatially localised, and provided the atoms are elastically relaxed then the configuration is metastable.
            We describe our okMC state as the types, and positions of objects that it contains, where the type of each object is stored as a hash key to an atomic configuration stored in a database.
            Note that this is a different approach to previous akMC-okMC hybrid techniques where a handshaking is made between atomistic and idealised objects\cite{Castin_JCP2011,Pannier_Thesis2017}- all objects here are fundamentally atomistic. This allows us to handle arbitrarily complex defects, including interstitial-type defects and complex combined vacancy-interstitial objects.
            
            In okMC the rules for evolving the state are often simplified to rigid body translations; a transformation from one object to another; and splitting one object into two.
            \footnote{The reverse event (coalescence), which combines two objects into one, is required by detailed balance and may occur at a much higher rate than dissociation. But this event is a result of object proximity rather than a fundamental dynamic process of one object.}
            These rules often are given as simple rates which vary depending on object type, but with atomistic detail it is possible to find these rates explicitly.
            The rates for evolving the atomistic defect objects are constructed on-the-fly by considering possible atomistic processes which change the configuration.
            
            Finally we note that in okMC the host crystal is hardly referenced, except possibly as an homogeneous elastic medium, in contrast to MD or standard atomistic KMC.
            In our simulation the atoms in the ``perfect crystal" far from the objects is not stored.
            In common with existing okMC codes such as MMONCA\cite{MartinBragado_CPC2013}, the computational cost using our code for a simulation of one object contained in a one thousand atom system is therefore similar to the cost of one object in one million atom system.
            In contrast with other okMC codes, there is only one type of object in our work- an atomic configuration, with one set of rules for its dynamics as described below.

        \subsection{An atomistic object}

            The problem of identifying an atomic configuration or transition are related and have been tackled previously with graph theory \cite{Beland_PRE2011} and bitmaps \cite{Nandipati_JPCM2016}.
            We will use the bit-twiddling Zobrist hashing method \cite{Mason_CPC2005}.
            We start with the observation that atoms repel each other at short range and so are never too close together.
            Consider a simple cubic crystal mesh with side $a_0/4$.
            The furthest apart two atoms could get while mapping to the same node is $a_0 \sqrt{3}/4$.
            This is 50\% the nearest neighbour separation on a bcc lattice side $a_0$, or 61\% the nearest neighbour separation on an fcc lattice.
            If atoms would prefer to be on bcc or fcc lattice sites, then except during high energy collisions they will never get so close together they will map to the same $a_0/4$ simple cubic node.
            For practical purposes any complex atomic configuration may be represented by having zero or one atoms slightly displaced from $a_0/4$ simple cubic nodes.
            If interstitial alloys are considered the validity of the finer lattice may need to be verified, but the principle of non-multiply occupied sites on a sufficiently fine lattice remains clear: if we cannot guarantee zero or one atom per node spaced by $a_0/4$, it may yet be possible if the node spacing is $a_0/8$.
            The extension to non-cubic lattices is similarly trivial.

            It is less clear that the mapping of the real space positions in $\mathcal{R}^{3N}$ to fine-scale nodes is one-to-one rather than many-to-one.
            For the former to be true, then only one relaxed configuration of atoms maps to a particular set of fine-scale nodes, and a small displacement of the atoms would relax back to the same point.
            This is impossible to guarantee for all potentials, as the watershed hypersurfaces between local minima can be arbitrarily complex \cite{Pickard_JPCM2011}.
            If during a kMC simulation, two slightly perturbed configurations of atoms were associated with a single local minimum, then the damage done to the dynamics would be small, as all the evolution is quasi-static and the barrier between such close minima is most likely negligible anyway.
            We therefore assert that each configuration of atoms on a fine mesh is associated with a unique minimum, and use the fine mesh occupations to define the configuration.
            This reduces the description of the configuration to the atomic-chess-board problem, and so we use the extremely efficient Zobrist key\cite{Mason_CPC2005} to find a 128 bit hash key for an atomic configuration.
            These hashes index stored objects - that is the atomic configurations plus their associated transitions - on a large database.
            We return to the issue of the size of the database needed in section \ref{MDcascades}.

            We define an object by first identifying all defected unit cells - meaning those which do not have the same occupation of the same fine-scale nodes as the perfect crystal.
            A buffer region of $m$ unit cells in each direction ( ie the cube of $(2m+1)^3$ cells including the defected unit cell ) is added, and then centered in a cubic minimal bounding box.
            This region of atoms is ascribed to an object.
            An object can therefore be rather large, spanning $32^3$ unit cells in some of our cascade evolution simulations, and contain both interstitials and vacancies.
            We have found that insisting on a buffer region of $m=2$ unit cells around defects is adequate to reproduce the elastic interaction energy between pairs of defects to within $10\%$, as determined by comparing the energy computed using the dipole tensor formalism (see section \ref{parameterization}) to a full atomistic relaxation.
            
            In this work we do not include alloying species.
            We would expect adding an interstitial or substitutional atom like hydrogen or helium to work using this same model for objects ( note that bcc tetrahedral $[ \half\,\quarter 0 ]$ and octahedral $[ \half\,\half 0 ]$ interstices are perfectly resolved ).
            As an aside, we note that alloying elements have previously been tackled using a grey-alloy approach \cite{Chiapetto_JNM2015}, but this is left for future work for our model.

        \subsection{Finding a transition}
            \label{transitions}

            In a lattice-based kMC simulation, a transition consists of a mobile atom species moving from one lattice site to another, or for a pair of atoms to exchange sites.
            The number of transitions is determined by the number of mobile atoms and the number of moves each can make.
            The rate of each transition may be found by consulting a table using the before and after local configurations.

            In an off-lattice kMC simulation, transitions can be sought on-the-fly using the dimer method\cite{Henkelman2002}, kART\cite{Beland_PRE2011,El-Mellouhi_PRB2008}, or AMD\cite{Perez_ARCC2009,Uberuaga_2007}.
            The number of transitions is limited by the length of computational time available to search, there will always be more high energy ( and therefore low probability ) transitions.
            Though it offers a more exhaustive list of possible transitions in principle, on-the-fly searching is not without its difficulties.
            Some transitions may be more readily found than others, and care must be taken not to double count them.
            It is also difficult to ensure that all relevant barriers are added.
            As an example consider the eight $\langle \half\,\half\,\half \rangle$ nearest-neighbour vacancy-atom exchanges in a perfect bcc lattice ( we ignore second neighbour $\langle 1 0 0 \rangle$ exchanges for clarity of exposition ).
            Each saddle has the same probability of discovery, so the number of times each saddle is found should be Poisson distributed.
            The probability of not finding a particular saddle is $\bar{p} = exp(-m/8)$, where $m/8$ is the mean number of hits per saddle given $m$ trials.
            The probability of finding all saddles is $p_{all} = (1-\bar{p})^8$.
            To have a $p_{all}>50\%$ chance of finding all 8 saddles, we need $\bar{p}<0.083$, and so $m>20$.
            To have a $p_{all}>99\%$ chance of finding all 8 saddles, we need $m>53$.
            If objects are to be reused, we need a good coverage of saddles, and so this is potentially a lot of work.
            Using the symmetry of the system could help reduce wasted work, but in general cases the local surrounding environment will show little symmetry and so make it difficult to achieve great gains.
            SEAKMC\cite{Xu_JPCM2012} avoids this problem by sampling the saddle points rather than attempting an exhaustive survey.
            This is efficient, but requires careful tuning before starting a simulation to get correct residence times.
            It would not be appropriate for a model where barriers are reused.
            Recently progress has been made to minimise the impact of unsearched transitions using the AMD framework to find transitions\cite{Swinburne_PRM2018}.

            An alternative route for computing the transition energy for this off-lattice atom-vacancy exchange example is to recognise that the move remains identifiable even when atoms are displaced slightly from lattice sites.
            For $\langle 111 \rangle$ crowdions the moves are also well-known - there exists a Frenkel-Kontorova string pull in the $\langle 111 \rangle$ direction\cite{Braun_PR1998,Derlet_PRB2007}, and a rotation through the $\langle 110 \rangle$ dumbbell to $\langle 11\bar{1} \rangle$\cite{Derlet_PRB2007}.
            We can code these moves by searching for the correct atomic environment to permit the move, then moving atoms appropriately.
            The number of possible transitions is easily established, and moves which have the same final configuration clearly have different routes, so double- counting them is not a problem.
            We explicitly sacrifice the chance of finding unusual or unexpected transition paths in favour of a reusable and smaller, but complete, set of transitions.

\begin{figure*}
                    \centering
                    \includegraphics[width=.9\linewidth]{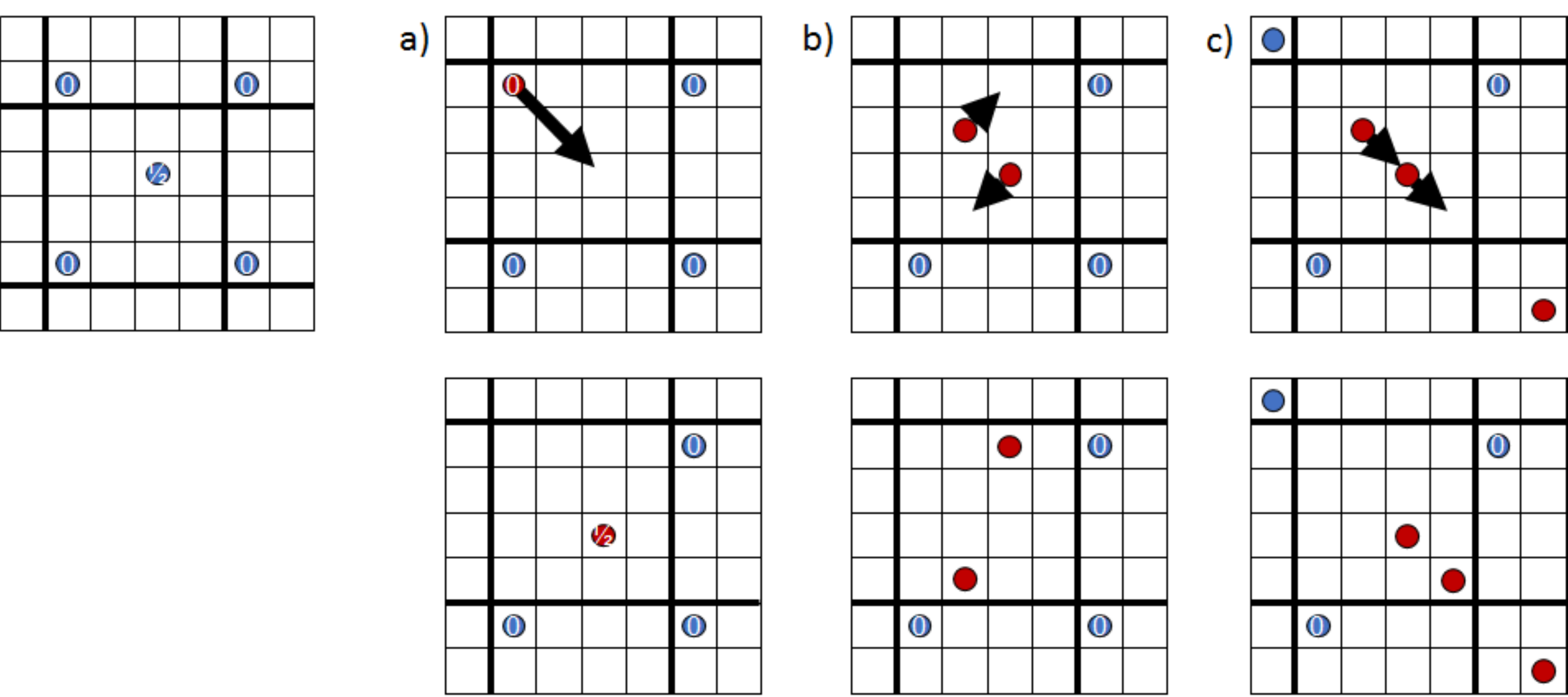}
               \caption{\label{fig:transitions}
                A cartoon illustrating transitions defined on a fine lattice, with the top row indicating the `before' and the bottom row the `after' configurations.
                From left to right, the bcc lattice, a) an atom-vacancy exchange, b) a rotation move, and c) a string pull move.
                Note that inside the code the atom positions are fully off-lattice, and the fine $a_0/4$ simple cubic lattice is present to distinguish quickly between similar states.
                }
            \end{figure*}

            With the fine mesh defined, we can define three prototype atomic transformations for the bcc lattice illustrated in figure \ref{fig:transitions}.
            \begin{description}
            	\item{Vacancy-atom exchange}:
            Find a high energy atom, and look for an unoccupied `vacancy' lattice site separated by $\langle \half\,\half\,\half \rangle$.
            We permit the transition attempt if the unoccupied site itself is surrounded by 26 unoccupied neighbours on the fine mesh lattice ( the region $-\quarter:\quarter$ ).

			\item{Rotation}:
            Look for a pair of high energy atoms, and displace in opposite $\langle \quarter\,\quarter\,\quarter \rangle$ directions.

			\item{String pull}:
            Look for a high energy atom, and then check in successive $\langle \quarter\,\quarter\,\quarter \rangle$ cells for two more high energy atoms.
            If they are present, then the original and its first neighbour are displaced along the string.
		\end{description}

%             We can consider motion of interstitial-like objects using the Frenkel-Kontorova (F-K) model ( for which a comprehensive review can be found in ref\cite{Braun_PR1998} ).
%             In this model, a $\langle 111 \rangle$ string of atoms is represented by its centre-of-mass, which moves over a (periodic) potential energy surface.
            We also need a fourth prototype transformation for the cooperative motion of clusters of atoms.
            The multi-string F-K model\cite{Dudarev_PM2003} is a simple extension for understanding the motion of clusters and loops.
            Here many $\langle 111 \rangle$ strings move over the potential surface, but are additionally coupled together with elastic springs.
            The springs represent the energy stored in the dislocation core, and acts to prevent one string getting too far ahead of its neighbours.
            If a single string of atoms is pulled over the PES, the restoring force of its neighbours will tend to bring it back again.
            There may not even be a relaxed metastable configuration of a dislocation line with just one string of atoms pulled.
            This is the physical origin of the Ohsawa transition\cite{Ohsawa_JNM2007} between motion of an interstitial cluster as a single rigid body and by the double-kink mechanism for a large dislocation loop.

            A good model might consider simultaneous movement of multiple strings of atoms.
            But it would be difficult to justify moving two strings together rather than three, or more.
            The multiplicity of final states to investigate would, at present, lead to an unsupportable slowdown of the code.

            For now we can ensure there is a macroscopic diffusion of the centre of mass of a cluster by exploiting the okMC nature of the model.
%            If an object is found to have many possible transitions which all relax back to the same original configuration, then it may be a candidate for cooperative centre-of-position motion.
            Within an object, we search for a contiguous subcluster of defected unit cells (excluding padding) containing only interstitials. 
            Then for this subcluster we check the principle eigenstresses, as computed by diagonalizing the dipole tensor.
            An interstitial cluster will have one large eigenstress along its Burgers vector.
            If these conditions are met, then the rate of transition to the same subcluster translated by $[\half\,\half\,\half]$ along the Burgers vector is fixed to $\nu/\sqrt{n_I}$, where $\nu$ is the attempt frequency of a single string pull ( see table \ref{tab:transitions} ) and $n_I$ the number of interstitials.
            This square root dependence of the prefactor has been observed in simulation\cite{Osetsky_PMA2000,Swinburne_PRL2014} and experiment\cite{Arakawa_Science2007}.
            We leave investigation of the cooperative motion of multiple strings to future work, and acknowledge that our model is currently limited to smaller interstitial clusters. 

            With these sets of moves and with the atoms off-lattice, a great range of atomic configurations can be explored.
            It is interesting to note, however, that all transition events that an object can make can be simply categorised.
            \begin{itemize}
                \item
            Events can define a transformation of the atomic configuration, taking one object to another.
                \item
            A transformation may require/permit a resizing of the object to preserve its buffer of perfect crystal.
                \item
            A transformation may translate the origin of the object.
            	\item
            Events can split the object into two or more.
            In our code we do not consider splitting into more than two objects- which can temporarily leave an object with parts far enough apart that they do not overlap.
            This is a rare case, and there is no damage done by this to the dynamic evolution anyway, only a minor suboptimal evolution until the full decomposition has been recognised.
            There is no requirement to include a transition event combining pairs of objects.
            This can be automatically included by testing for the overlap of objects at the beginning of every KMC step.
            \end{itemize}

        \subsection{Parameterization of kMC simulations}
            \label{parameterization}

            All kMC simulations reported here use the same parameterization, with no additional tuning.
            If the transition takes an object from configuration $(A)$ to $(A')$,
            the transition rate $(A \rightarrow A')$ is taken to be
                \begin{equation}
                    \label{eqn:rate}
                        r_{(A \rightarrow A')} = \nu_{(A \rightarrow A')} \exp\left[ - \frac{ \Delta E_m^{(A \rightarrow A')} }{k_B T} \right],
                \end{equation}           
            where $\nu_{(A \rightarrow A')}$ is a rate prefactor (which itself might be temperature-dependent) and $\Delta E_m^{(A \rightarrow A')}$ a temperature-independent migration barrier.        
            As shown in ref\cite{Mason_JPCM2017}, for pure tungsten at least, the simple Kang-Weinberg model\cite{Kang_JCP1989} is a good approximation for the migration barrier for an atom-vacancy exchange.
            If we consider a transition which takes an object configuration $(A)$ to $(A')$, then the migration barrier is
                \begin{equation}
                    \label{eqn:barrier}
                    \Delta E_m^{(A \rightarrow A')} = E^{(A \rightarrow A')}_{saddle} - E^{(A)} = \max\left\{  \frac{E^{(A')} - E^{(A)}}{2} + \Delta E_m, \frac{\Delta E_m}{2} \right\},
                \end{equation}
            where $\Delta E_m$ is a constant for the transition category, and $E^{(A)}$ is an appropriate ``relaxed" energy for object configuration $(A)$.
            $\Delta E_m$ barriers for all moves are given in table \ref{tab:transitions}.
            The maximum function is used to ensure that the migration energy is always positive; $E^{(A)}$ may take any value but $(A)$ is known to be metastable.
            We take the minimum value for the barrier as $\frac{\Delta E_m}{2}$, which is an empirical choice taken to be sufficiently low to be rarely needed in practice, and sp not unduly affect the dynamics.
            For fast-moving interstitial clusters the barrier is close to zero in any case; for vacancy defects this limit will only be applied where there is a strong driving force to select one move in preference to another.
            We define the appropriate relaxed energy below.

            As previously noted, we store atomic configurations of defects as localised objects surrounded by a buffer region of perfect crystal, and an object can be large.
%             It would be expensive to relax the whole object, and may not even be physically correct.
%             The maximum speed of information propagation should be the speed of sound- of order a few nanometres per picosecond.
            We compute the energy barrier for a transition by first finding the region where atoms are expected to move, then adding a buffer region to this ( see table \ref{tab:transitions} ).
            This cut-out region of 8-10 unit cells can be significantly smaller than the original object which might be 20-30 unit cells..
            We fix atoms in the outermost unit cell boundary of the cut-out, and relax.
            This takes the energy of the cutout region from $E_{cut}^{(A)}$ to $E_{cut}^{(A')}$.

            Fixing the atoms ensures that the transition region after the event fits back into its parent object, and takes some account of the strain field in which the transition is located, but note that this is \emph{not} treating the elastic interaction fully.
            Firstly, we need to account for the response of the remainder of the object to the transition.
            If $\vec{f}_{embed}$ is the force on the atoms in the object introduced by reintroducing the cutout after the event, and $D_{ab} = \partial^2 E / \partial \vec{r}_a \partial \vec{r}_b$ is the Hessian computed with atoms in their `after' positions, then the linear elastic response in the object is
                \begin{equation}
                    E_{embed} = - \half \vec{f}_{embed} \cdot D^{-1} \vec{f}_{embed}.
                \end{equation}
            This can be evaluated using a Lanczos recursion technique\cite{Mason_JPCM2004,Rudd_PMS2007}, and typically takes order ten milliseconds on a single core, a few percent of the time taken to fully relax the atom positions in the cut-out region.
            Secondly, as the object is defined by atomic positions, we need to account for the self-energy of the object due to its periodic images.
            This can be done using the method of Varvenne et al\cite{Varvenne_PRB2013}.
            As we are using embedded atom potentials, the dipole tensor is easily computed using
            \begin{equation}
                P_{ij}  = - \sum_a \sum_{b \in \mathcal{N}_a} \left( {^1\!/\!_2} \left. \frac{ \partial V}{\partial r} \right|_{r_{ab}} + \left. \frac{ \partial F}{\partial \rho} \right|_{\rho_a} \left. \frac{ \partial \phi}{\partial r} \right|_{r_{ab}} \right)
                    \frac{ r_{ab,i} r_{ab,j} }{r_{ab}},
            \end{equation}
            where $r_{ab,i}$ is the $i^{th}$ Cartesian coordinate of the separation between atoms $a$ and $b$, $V(r)$ is a pairwise potential and $F[\rho]$ a many body contribution from the embedding density function $\rho=\sum \phi(r)$.
            This is a simple sum over atoms and their neighbours, using the same first derivatives as a force, and so is of negligible computational cost.
            The elastic energy between an object with dipole tensor $P^{(A)}_{ij}$ and a second with dipole tensor $P^{(B)}_{ij}$ located at separation $\vec{R}_{AB}$ is\cite{Dudarev_Acta2017}
                \begin{equation}
                    \label{elasticEnergyDipoleTensor}
                    E_{elas}^{(A,B)}\left(\vec{R}_{AB}\right) = P^{(A)}_{ij} \frac{\partial^2 G_{ik} \left(\vec{R}_{AB}\right)} {\partial x_j \partial x_l } P^{(B)}_{kl},
                \end{equation}
            where $G_{ij}$ is the elastic Greens function. 
            In this work we use the isotropic elastic Green's function for convenience, as tungsten is nearly elastically isotropic.
            This requires the bulk elastic Lam\'e parameters $\lambda$ and $\mu$, which we compute at the beginning of the simulation using the empirical potential supplied.
            The centre-of-position of an object is taken to be \cite{Dudarev_NF2018}
            \begin{equation}
                \label{centreOfPosition1}
                {\bf R} = \frac{ \sum_{a=1}^{n}\,\left\lVert {\bf P}^{a} \right\rVert {\bf R}^{a} }{\sum_{a=1}^{n}\,\left\lVert {\bf P}^{a} \right\rVert  }
            \end{equation}
            where $\left\lVert {\bf P}^{a} \right\rVert = \sqrt{ \mathrm{Tr}( ({\bf P}^{a})^2 ) }$ is the Frobenius norm- a measure of the strength of the stress field generated by the $a^{th}$ atom.

            The self energy of the object is found in principle by the conditionally convergent sum of equation \ref{elasticEnergyDipoleTensor} over defect periodic images\cite{Varvenne_PRB2013,Dudarev_PRM2018}:
                \begin{equation}
                    \label{selfEnergy}
                    E_{self}^{(A)} = \sum_{uvw} E_{elas}^{(A,A)}\left(\vec{R}_{uvw}\right),
                \end{equation}
            where $\vec{R}_{uvw}$ is the position of the image translated $\{u,v,w\}$ periodic repeats in each Cartesian direction.
            To avoid convergence issues we simplify using $\{u,v,w\} \in [-1:1]$.

\begin{figure*}
                    \centering
                    \includegraphics[width=.9\linewidth]{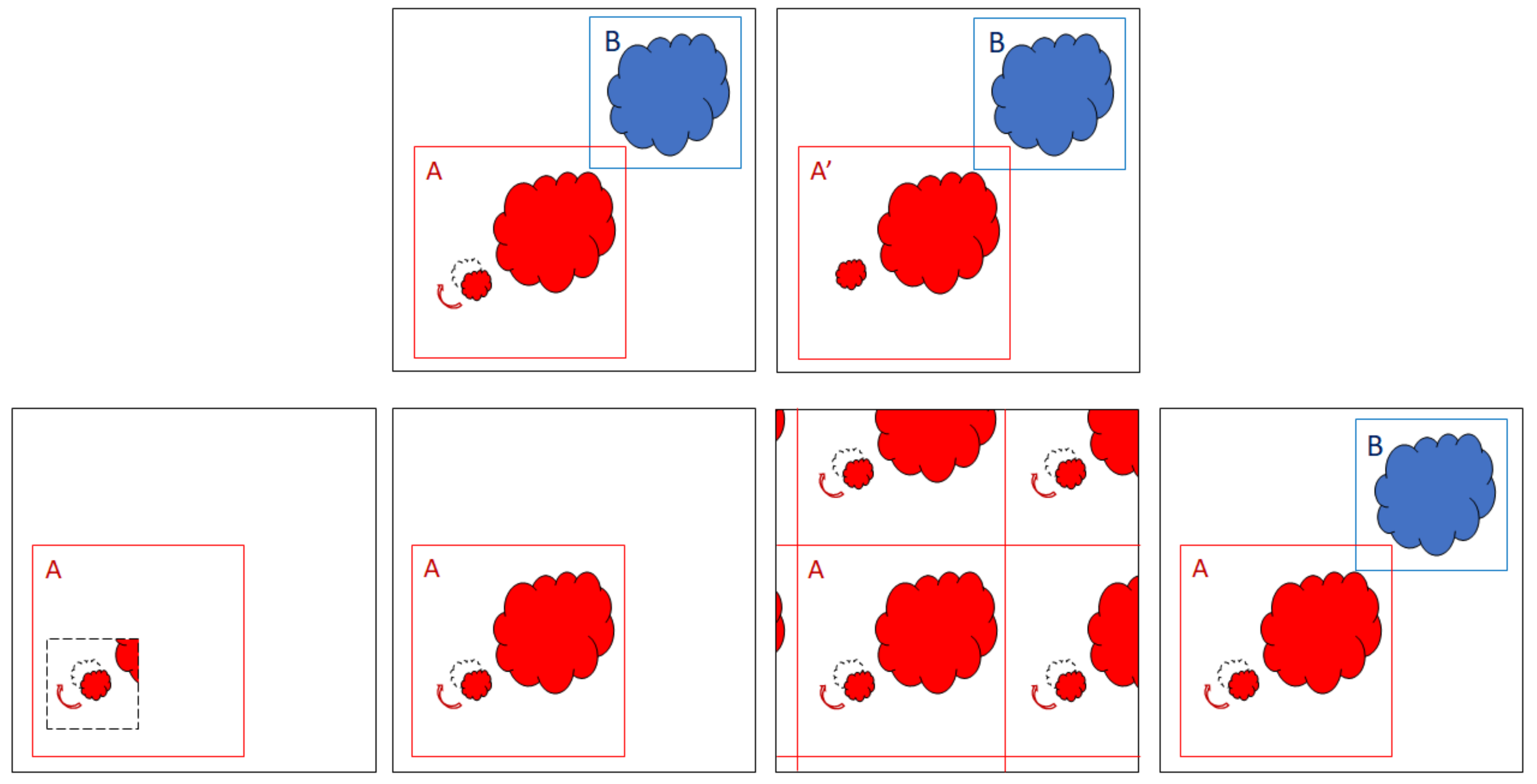}
               \caption{\label{fig:relaxation}
                A cartoon illustrating a transition $(A)\rightarrow(A')$.
                The top row shows a small rearrangement of the atoms in object $(A)$.
                To compute the rate of the transition ( equation \ref{eqn:rate} ), the steps are indicated in the bottom row as follows:
                i. A small box surrounding the moving atoms is defined, indicated by the hashed lines, with the atoms on the boundary fixed.
                The energy of the atoms in the cutout region is computed using empirical potentials.
                ii. The relaxation of the object due to re-embedding the cutout is computed at the harmonic level using the Hessian over the whole object $(A')$.
                iii. The self-energy of the object due to interactions with periodic copies of itself is computed using its dipole tensor.
                iv. Finally the interaction of the object with a distant second object $(B)$ is computed with dipole tensors.
                These four energy differences are inserted into equation \ref{eqn:barrier}.
                Finally table \ref{tab:transitions} is consulted to find the rate prefactor and fixed component of the energy barrier.
                }
            \end{figure*}

            Thirdly we need to account for the changing elastic response of the system as a whole.
            This is again done with the dipole tensor formalism, an approach previously exploited by Subramanian et al\cite{Subramanian_PRB2013}, and in a simplified form in ref \cite{Mason_JPCM2014}.
            If the transition takes the object from configuration $(A)$ to $(A')$, then the energy difference is
                \begin{eqnarray}
                    \label{state_energy_difference}
                        E^{(A')} - E^{(A)} &=& E_{cut}^{(A')} - E_{cut}^{(A)}          \nonumber \\
                                           &+& E_{embed}^{(A')} - E_{embed}^{(A)}       \nonumber \\
                                           &+& E_{self}^{(A')} - E_{self}^{(A)}         \nonumber \\
                                           &+& \sum_{B\neq A} \left( P^{(A')}_{ij} \frac{\partial^2 G_{ik} \left(\vec{R}_{A'B}\right)} {\partial x_j \partial x_l }
                                              - P^{(A)}_{ij} \frac{\partial^2 G_{ik} \left(\vec{R}_{AB}\right)} {\partial x_j \partial x_l } \right) P^{(B)}_{kl}.
                \end{eqnarray}
            The spatial regions for each level of relaxation are illustrated in figure \ref{fig:relaxation}.

            The migration barriers, characteristic frequencies and padding using for the cutout region are given in table \ref{tab:transitions}.
            As a side-benefit of having categories of transition defined, the rate prefactor may itself be a function of temperature or characteristic of the defect structure. 
            We set the rate prefactor of string-pull type low-energy barrier moves to be linearly dependent on temperature, corresponding to the friction limited regime \cite{Fitzgerald_PRL2008}.
            
            Note that the embedding energies can be stored as data for the transition and reused , and the self-energy is a property of the object.
            The long-range inter-object energy of interaction needs to be recomputed at every okMC step.
            If the dipole tensor is stored on the object, this takes negligible time compared to relaxing the atoms.

    \begin{table*}
        \centering
        \begin{tabular}{l|llll}
            move            &   high energy     &   barrier     &      rate             &   padding     \\
                            &   atom (eV)       &   $\Delta E_m$ (eV)        &      prefactor (THz)  &   (u.c.)      \\
            \hline
            atom-vacancy    &   0.25            &   1.75        &      6.45             &   2           \\
            exchange        &                   &               &                       &               \\
            \hline
            atom pair       &   1.10            &   0.40        &      1.0             &  3           \\
            rotation        &                   &               &                       &               \\
            \hline
            string pull     &   0.50            &   0.013       &      0.75 $k_B$T       &   3        \\
            \hline
            cooperative     &   cluster         &   0.013       &      0.75 $k_B$T / $\sqrt{N}$      &    2    \\
       \end{tabular}
        \caption{\label{tab:transitions}
        Parameterization of the three akMC and one okMC transition events considered in our simulations.
        High energy atoms with potential energy over the threshold are considered as candidates for the move.
        The migration barriers in equation \ref{eqn:barrier} use these tabulated values, weighted by the difference in before- and after- energies, according to equation \ref{state_energy_difference}.
        The rate $6.45$ THz is the Debye frequency for tungsten \cite{Becquart_JNM2010}.
        The energy barrier for vacancy migration is taken from DFT \cite{Nguyen-Manh_JNM2007}, which is in good agreement with experiment \cite{Mundy_PRB1978}.
        The energy barrier for string pull and crowdion rotation are taken from DFT \cite{Derlet_PRB2007}.
        The string pull and translation moves are assumed to be friction limited, with rate constant linearly proportional to $k_B T$.
            }
    \end{table*}

    We compute and store all the transition rates ( according to our search rules ) and so can use rejection-free kinetic Monte Carlo\cite{Bortz_JCP1975} using the basin-autoclimbing Mean Rate Method\cite{Novotny_PRL1995}.

    \section{Results}

        \subsection{Individual defects}
            \label{individualDefects}

            Before moving to complex systems with multiple moving defects, we present some validation work with isolated individual defects.
            This offers a comparison with literature, and demonstrates the working of the code.

            We consider the dynamics of a single monovacancy, a quad-vacancy cluster, a monointerstitial, and interstitial clusters of size 2,7,13,19.
            Here, all are simulated at 600K, using an empirical potential known to be good for vacancy defects ref\cite{Mason_JPCM2017}.
            In section \ref{intClusters} we show this empirical potential is also good for interstitial defects.

\begin{figure*}
              \includegraphics[width=.5\linewidth]{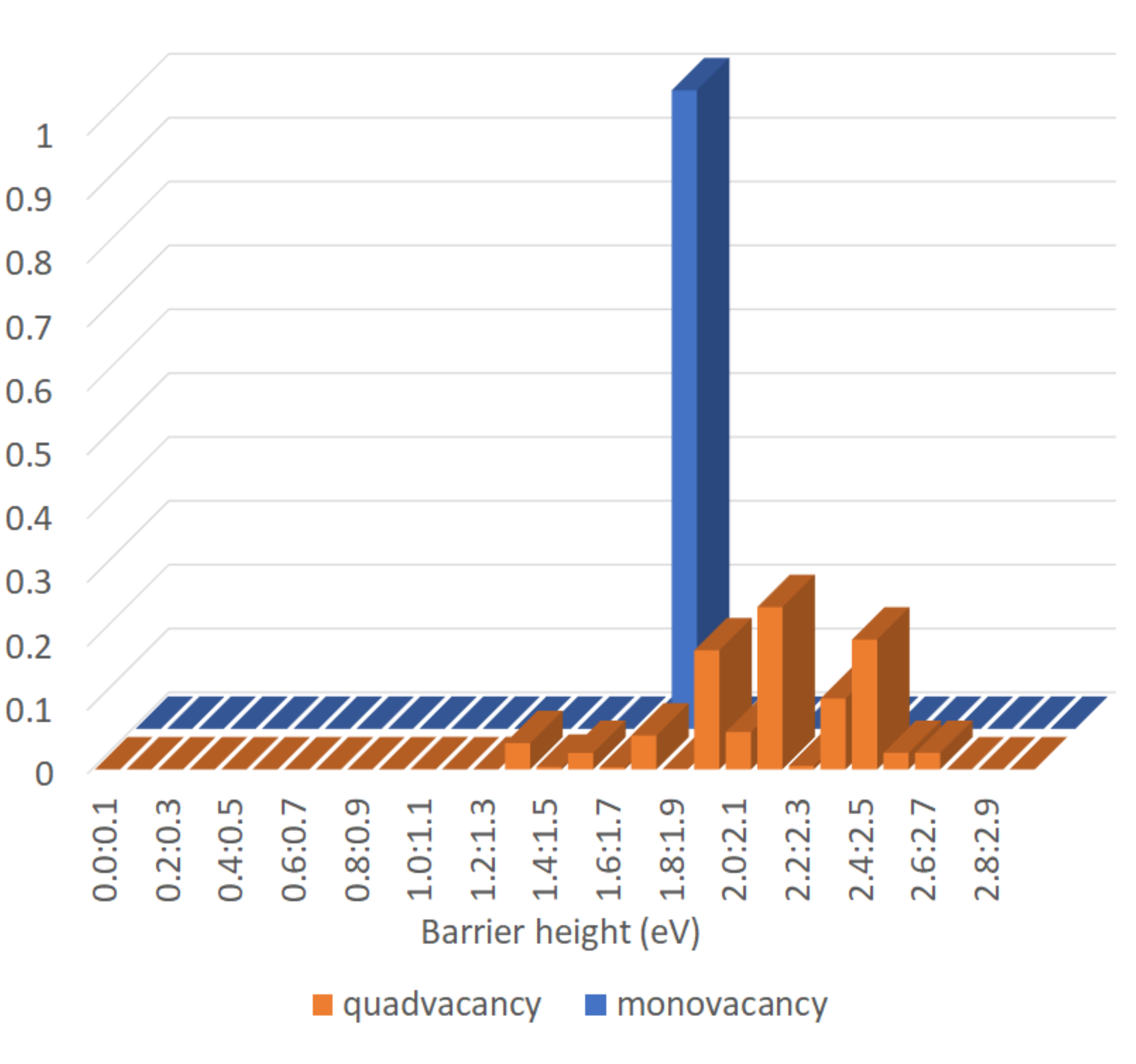}
              \includegraphics[width=.5\linewidth]{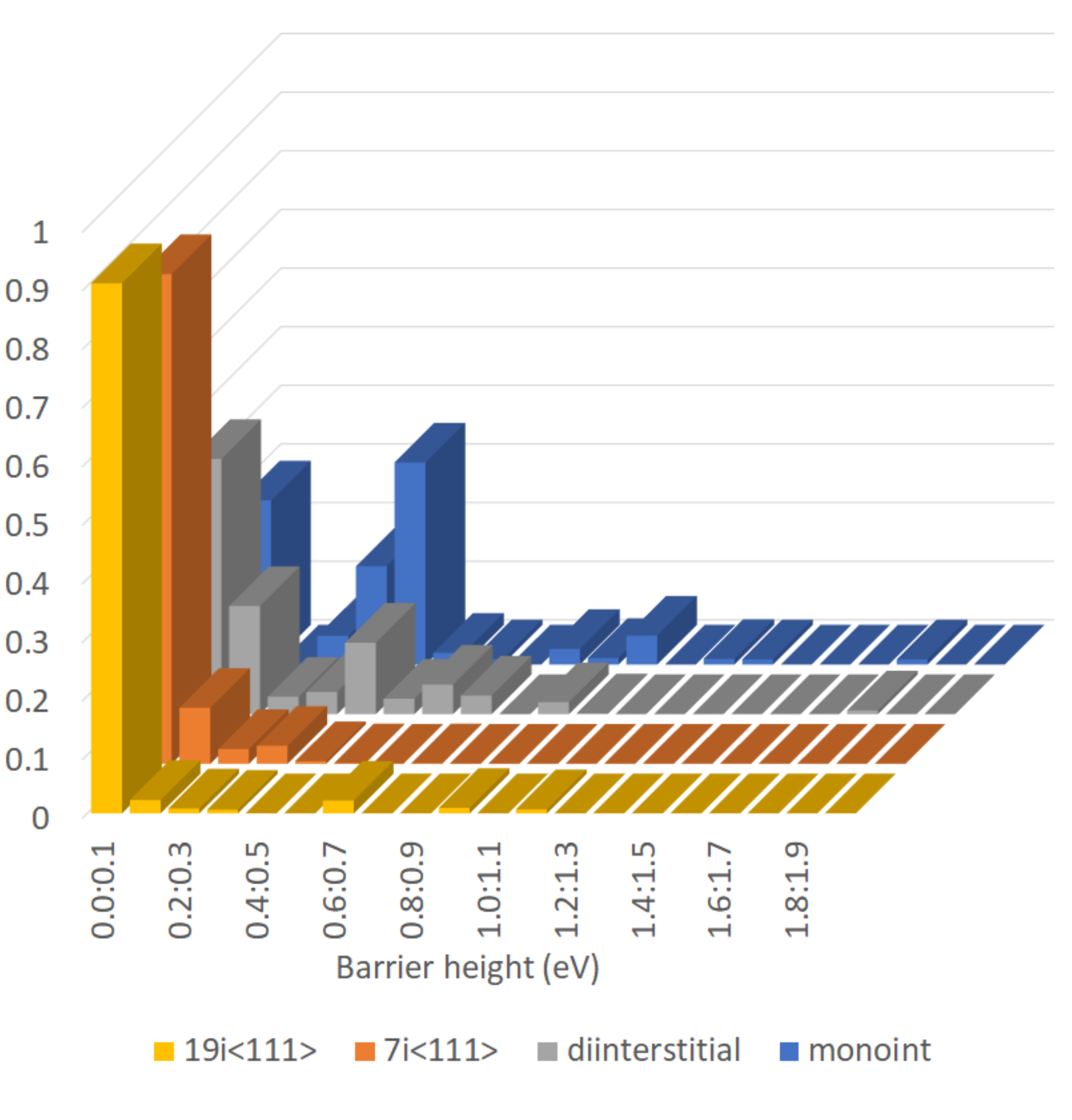}
               \caption{\label{fig:barrier_hist}
               Histograms of relative frequencies of migration barriers computed over 1000 kMC steps at 600K, using the MNB potential in tungsten.
               Top: vacancy-type objects. A single monovacancy has a barriers at exactly 1.75eV by construction, but a quad-vacancy explores a wider range.
               As the quad-vacancy itself is a low-energy cluster, many are above the isolated vacancy barrier.
               Bottom: interstitial-type objects. A single crowdion shows barriers near 0eV and 0.4eV, corresponding to translation and rotation modes, and a few higher energy transitions to other single-interstitial formations, octahedral, $\langle 100 \rangle$ dumbbell etc.
                The diinterstitial has a wider range of translation modes where one string pulls past the other, and a few rotations are found.
                The 7 interstitial cluster and 19-interstitial loop show mostly translation modes.
                }
            \end{figure*}

            In section \ref{parameterization} we explained how the migration barrier $\Delta E^{(A \rightarrow A')}$ was computed.
            In figure \ref{fig:barrier_hist} we present histograms of barriers computed over a selection of 1000 step simulations.
            The monovacancy always has equal-energy before- and after- states, so every barrier is exactly $\Delta E^{(A \rightarrow A')} =\Delta E_m = 1.75$ eV.
            By contrast the quadvacancy is a low energy cluster, so to evolve it must gain energy.
            It therefore shows a range of barriers above $1.75$ eV, and a few transitions back to the low-energy cluster below $1.75$ eV.
            The crowdion shows translations and transformations between crowdion and dumbbell near 0.0 eV, and some rotations near 0.4 eV.
            It also shows some high energy barriers to higher energy single-interstitial formations.
            The interstitial clusters show an increasing fraction of low-energy string-pull and cooperative cluster transitions either to different configurations or translations.

\begin{figure*}
                \includegraphics[width=.9\linewidth]{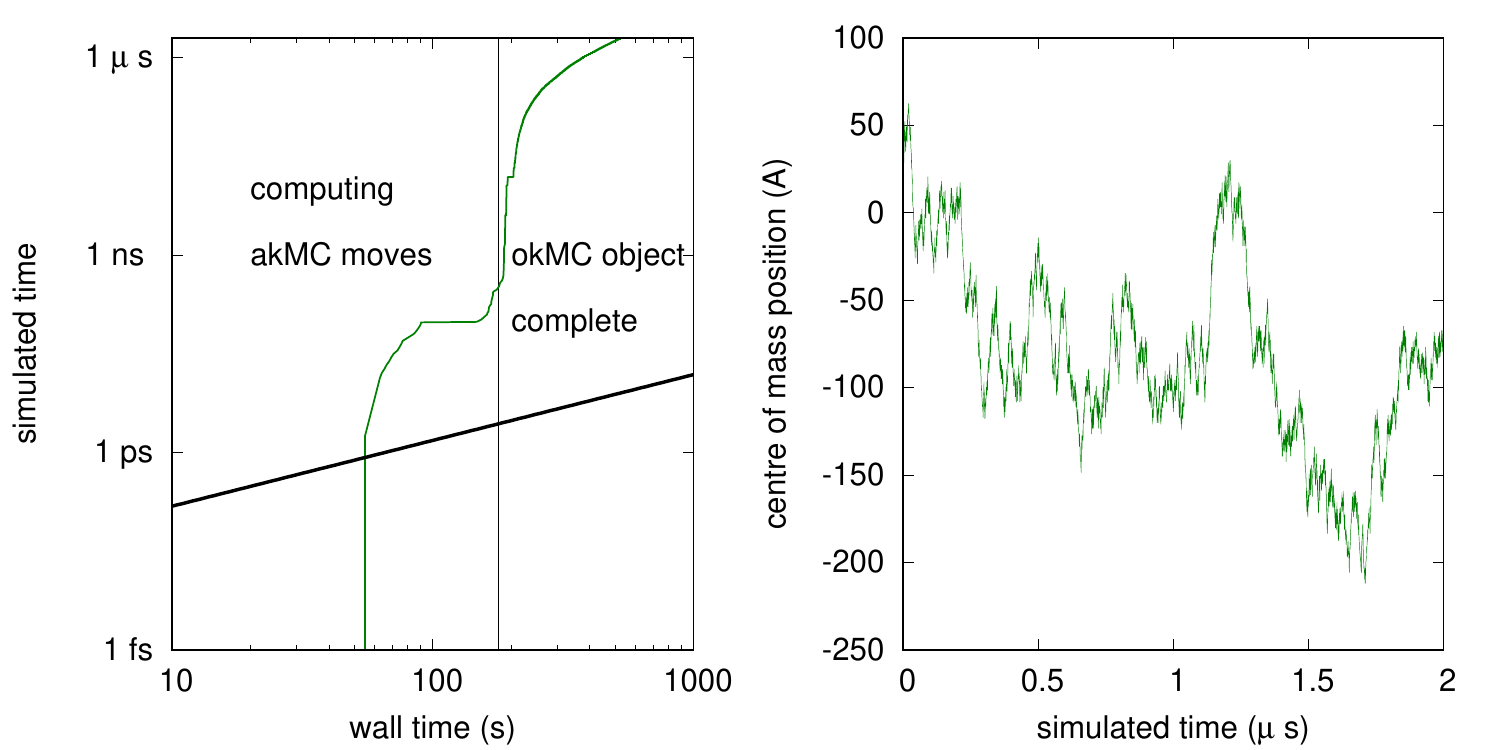}
               \caption{\label{fig:iclust}
               Data from a single run computing the diffusion constant of a 13 interstitial cluster at 600K in a $32\times 32\times 32$ unit cell box.
               Left: for three minutes wall time ( using a single core on a desktop PC ), string pulling moves are considered and stored. After this they can just be recalled, and so the code accelerates.
               The solid black line is an indication of MD speed, assuming a good MD code can perform one million atom update steps per second (per core), and each timestep is 1fs.
               Right: the (x-) position of the centre of mass of the cluster, computed using equation \ref{centreOfPosition1}, showing expected Brownian motion with no significant drift.
               This is a trivial exercise using standard okMC where the rules for translations are predetermined, but less so for an on-the-fly code where the rules for translation are computed.
                }
            \end{figure*}

            When all low energy barriers describing translations and transformations have been computed, the object KMC model is fully determined, and so the code can evolve them at the same speed as any other okMC.
            This is illustrated in figure \ref{fig:iclust}.
            On this plot is also indicated the performance of a good MD code such as LAMMPS\cite{LAMMPS}, which currently takes about one second wall time to perform an update step for 1 million atoms, the update step being a simulated time of one femtosecond.
            An important point to note in figure \ref{fig:iclust} is that our KMC code starts slower, but becomes faster than this canonical benchmark for MD because of the efficient reuse of object information.

\begin{figure*}
                \includegraphics[width=.9\linewidth]{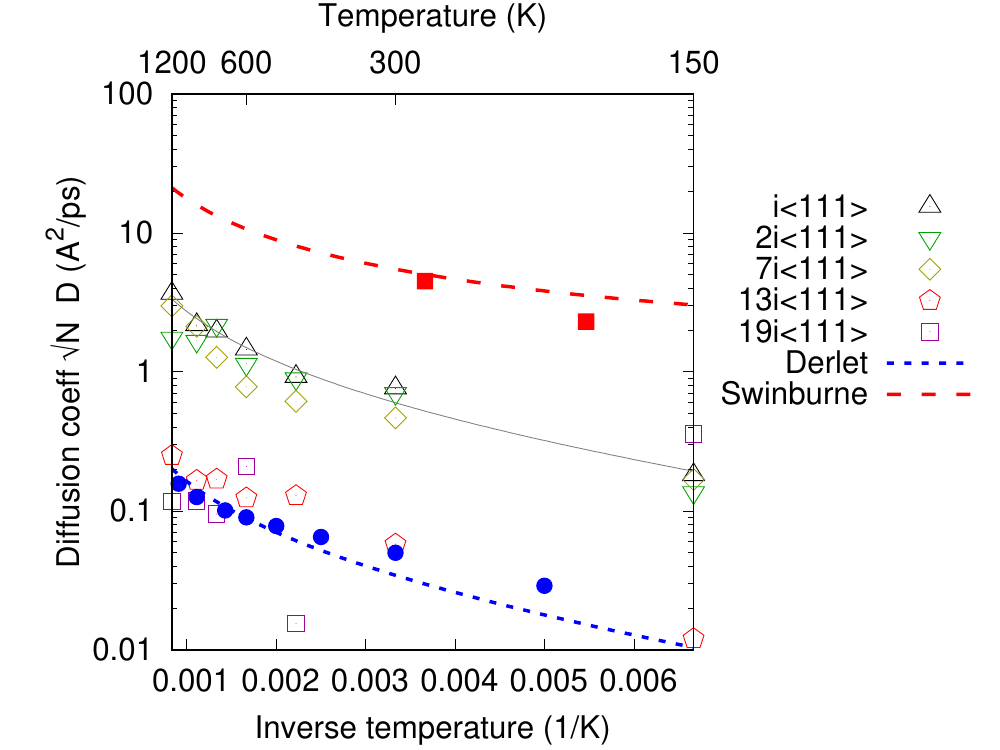}
               \caption{\label{fig:diff}
               The diffusion constant ( points ) computed using our model, multiplied by the square root of the interstitial count.
               The dashed lines are the theoretical prediction from Swinburne et al\cite{Swinburne_NJP2017}, and Derlet et al \cite{Derlet_PRB2007}, and solid symbols data points from MD simulations reported in those papers.   
               We conclude that the model presented here has order-of-magnitude correct diffusion coefficients for isolated interstitial clusters.
               Note that non-Arrhenius behaviour can be seen in this plot, a consequence of the temperature-dependent rate prefactor ( see table \ref{tab:transitions}. )
                }
            \end{figure*}

            We have computed the diffusion constant for interstitial defects using the method of ref\cite{Derlet_PRB2007}.
            The result is shown in figure \ref{fig:diff}.
            The string-pull attempt frequency and energy barrier are fixed to give a good crowdion diffusion constant.
            The computed diffusion constant for clusters is within an order of magnitude of literature results.
            This error is not expected to have a significant impact - clusters undergo a fast macroscopic displacement of the centre of mass, so this will not be a rate limiting step in longer term microstructural evolution.
            We would prefer not to overfit the model by introducing additional rules.

        \subsection{kMC annealing of primary radiation damage cascades}
            \label{MDcascades}
            
            In this section we anneal the complex defects produced in MD simulations of irradiation damage cascades.

            Irradiation cascade simulations were performed using the classical molecular dynamics code PARCAS \cite{PARCAS}.
            The simulations were performed in bulk tungsten, using the interatomic potential for tungsten by Derlet et al. \cite{Derlet_PRB2007}, stiffened at short range\cite{Bjorkas_NIMB2007}.
            Simulation cells were $68\times 68\times 68$ unit cells ( 629k atoms ), initially at 0 K, with the atoms on the periodic boundaries thermostatted to 0 K\cite{Berendsen_JCP1984}.
            Electronic stopping was modelled using a frictional force applied to atoms with kinetic energy over 10 eV, with the magnitude of the electronic friction determined by SRIM \cite{SRIM}.
            As discussed in section \ref{parameterization}, our model is currently only parameterized to correctly simulate the dynamics of small interstitial clusters, so we use low energy cascades, which have a low probability of producing a large loop \cite{DeBacker_EPL2016,Sand_EPL2016}.
            One atom was given an initial kinetic energy of 20 keV, and the simulations followed until cool ( $<10$ K ) at 40 ps.
            Details of the simulation method are given in \cite{Sand_JNM2014}.
            Five cascade configurations so formed were used as the starting point for further simulations described below.
            We cut out a cubic region of $40\times 40\times 40$ unit cells ( 128000 atoms ) containing the cascade with ample buffer.
            The simulations were run at 300K,600K and 900K multiple times with different random seeds.
            
            For these cascade evolution simulations it was necessary to use a large database of 30000 stored objects and 400 visited states, with the most recent 20 states used to construct the equilibrating basin \cite{Novotny_PRL1995}.
            Over the course of the simulation many more than 30000 different objects- many only separated by minor atomic configuration changes- were visited. 
            To prevent the database from continually growing, the next pensionable state replacement strategy\cite{Mason_CPC2005} was used to purge the database of less recently used objects.
            The memory footprint of each simulation was therefore large- about 12Gb- but constant.

            One additional simulation technique used for the cascade simulations is to employ absorbing boundary conditions.
            This is simple to implement in atomistic KMC: at the beginning of each move all atoms in the unit cells on the boundary of the simulation are set to be perfect crystal. 
            If this affects an object, then the object is relaxed before continuing, but now it may contain a different atom count.
            Using this method there is no strong bias towards the boundary, but neither will the defect return from the boundary.

\begin{figure*}
                    \centering
                    \includegraphics[width=.9\linewidth]{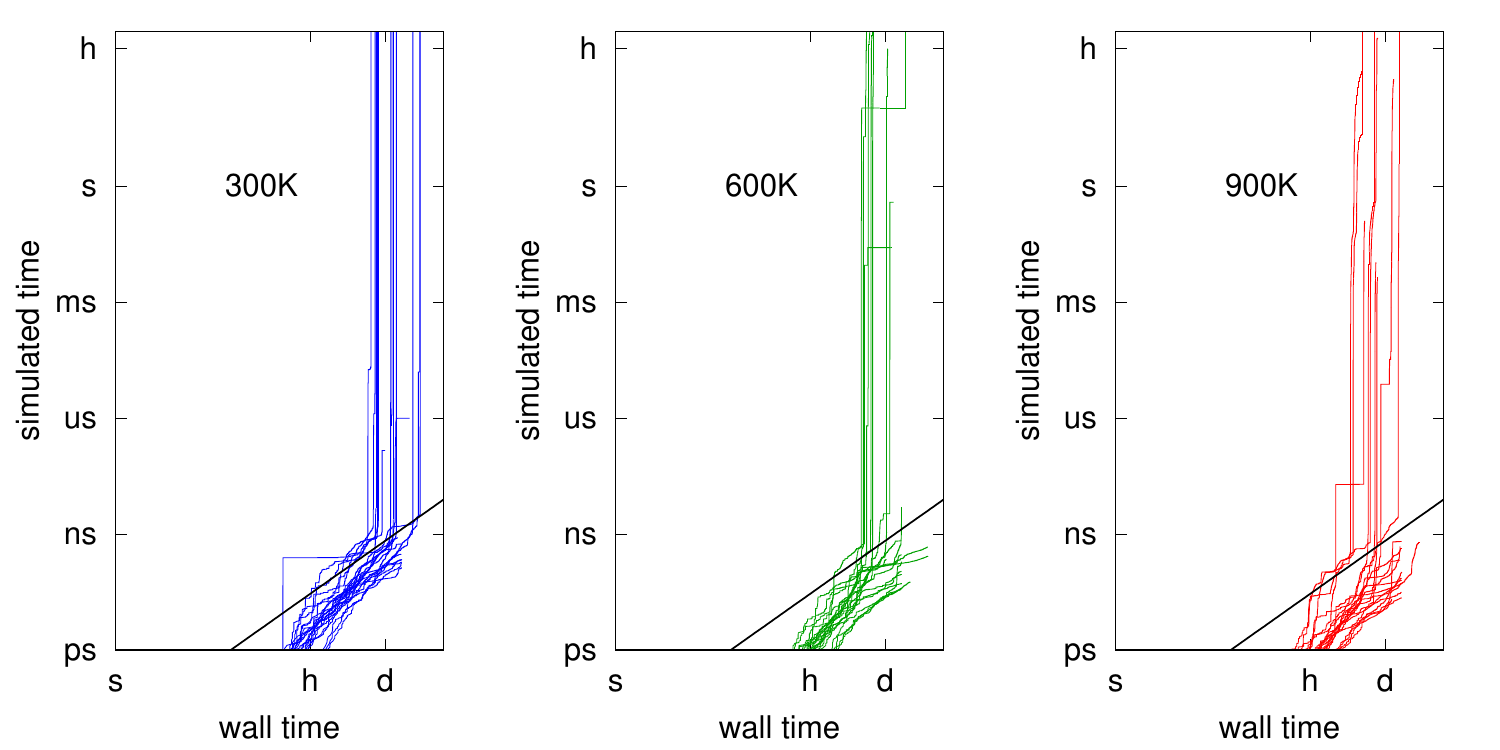}
               \caption{\label{fig:move_stats}
                Performance of the code, plotted as simulated vs wall time.
                The thick black line corresponds to a good MD code, running at 1 fs/1M atoms/s.
                After the first nanosecond of simulated time the KMC code overtakes MD.
                Note that we are considering multiple independent simulations on a single processor, and that we acknowledge parallelizing MD is significantly easier than parallelizing kMC.
                }
            \end{figure*}

            First we consider the timing results of our KMC code. 
            It is important that an off-lattice KMC code is actually faster than MD, and this is a difficult milestone to reach.
            We see in figure \ref{fig:move_stats} that our code overtakes MD after about 1ns of simulated time, which corresponds to about 24 hours wall time using a single processor.

\begin{figure*}
                    \centering
                    \includegraphics[width=.9\linewidth]{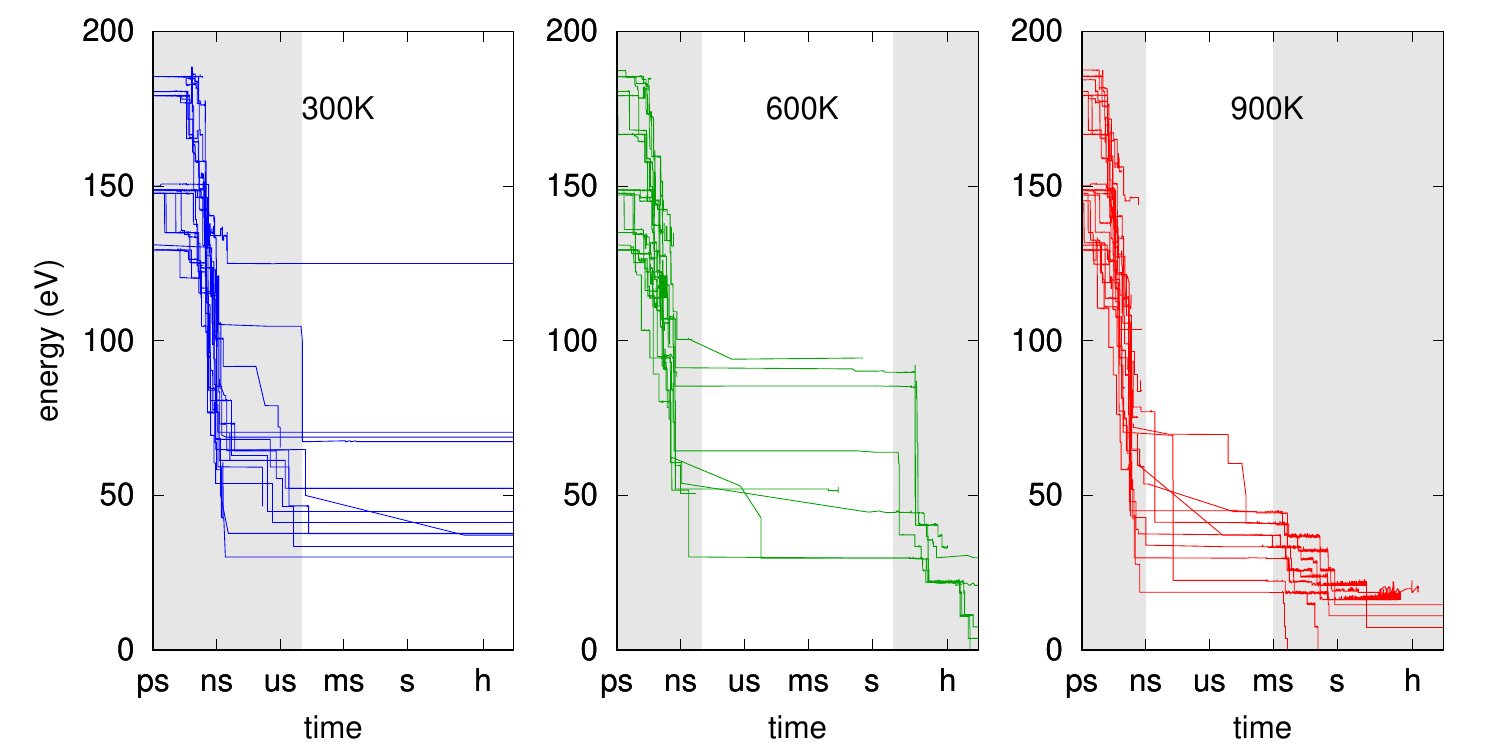}
               \caption{\label{fig:energy_stats}
                Evolution of the energy in the cascades considered as a function of simulated time.
                We can broadly split the evolution into the three parts, indicated by the vertical bands. 
                The greatest reduction in energy comes in the first few ns of simulated time.
                After this the energy evolves through a number of descending plateaux.
                At the longest times simulated vacancy motion becomes possible, and further relaxation is possible.
                }
            \end{figure*}

\begin{figure*}
                    \centering
                    \includegraphics[width=.9\linewidth]{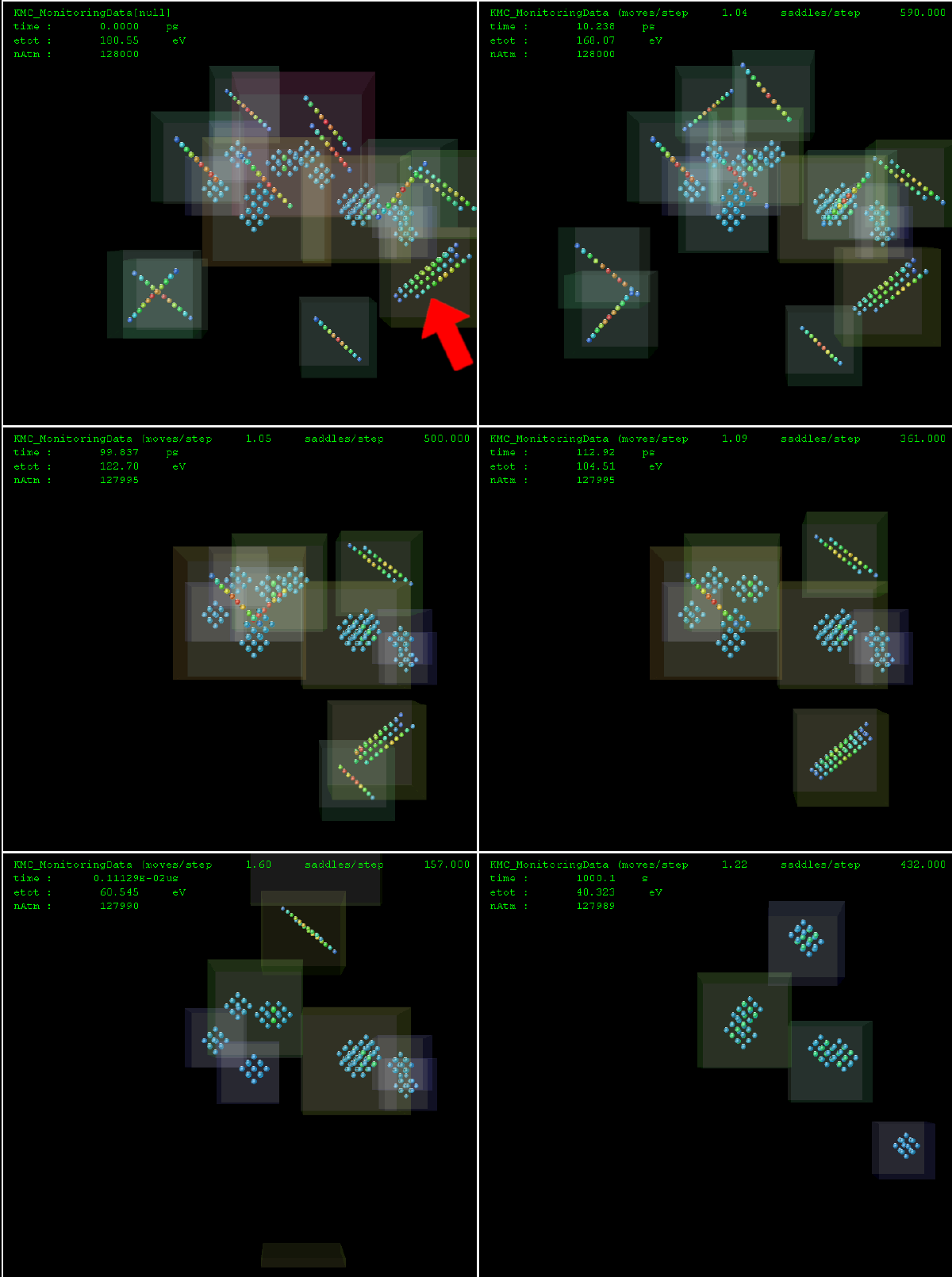}
               \caption{\label{fig:aokMC2}
               Snapshots from the evolution of a 20keV irradiation damage cascade in tungsten evolving at 600K.
               Reading left-to-right and top-to-bottom, the snapshots are taken at time $t=0, 10ps, 100ps, 113ps, 100ns, 1000s$.
               Starting from the top left image, the cascade starts with a number of crowdions.
               After 100ps three have diffused out of the simulation cell, and two more have annihilated on vacancies.
               At the bottom right we see a loose $4i\langle 111 \rangle$ interstitial cluster, indicated by a red arrow, gather a crowdion to become a $5i\langle 111 \rangle$ interstitial cluster.
               After 150ps only two small interstitial clusters remain, and by 120ns both have left the simulation cell.
               Brownian motion of the vacancies ultimately leads to accumulation into small 3- and 4- vacancy clusters.
               Only high energy atoms are shown, coloured from blue ( $+0.1$eV ) through green to red ( $+1.5$eV ).
               This calculation took 26 h on a single processor.
                }
            \end{figure*}

            Next we consider the evolution of the energy of the system.
            In figure \ref{fig:energy_stats} we see that the greatest portion of the energy reduction is in the first nanosecond of simulated time.
            This again we attribute to crowdion diffusion, absorption and annihilation.
            Figure \ref{fig:aokMC2} shows this behaviour for a cascade which starts with a diffuse collection of small defect clusters.
            Three crowdions are lost to the boundaries within 100 ps and two more have annihilated by recombination with vacancies.
            Two further crowdions are elastically trapped by each other and a small cluster of vacancies nearby.
            After 150 ps only two interstitial clusters remain, and both are glissile, so that after 120 ns both have left the simulation cell.

            After one nanosecond simulated time, the system evolves through a set of descending energy plateaux.
            This is common behaviour in kinetic Monte Carlo, associated with escaping trapping basins of energy states.
            In our case we find sessile interstitial clusters form through collisions of smaller clusters. 
            These then eventually relax to mobile clusters.
            
\begin{figure*}
                    \centering
                    \includegraphics[width=.9\linewidth]{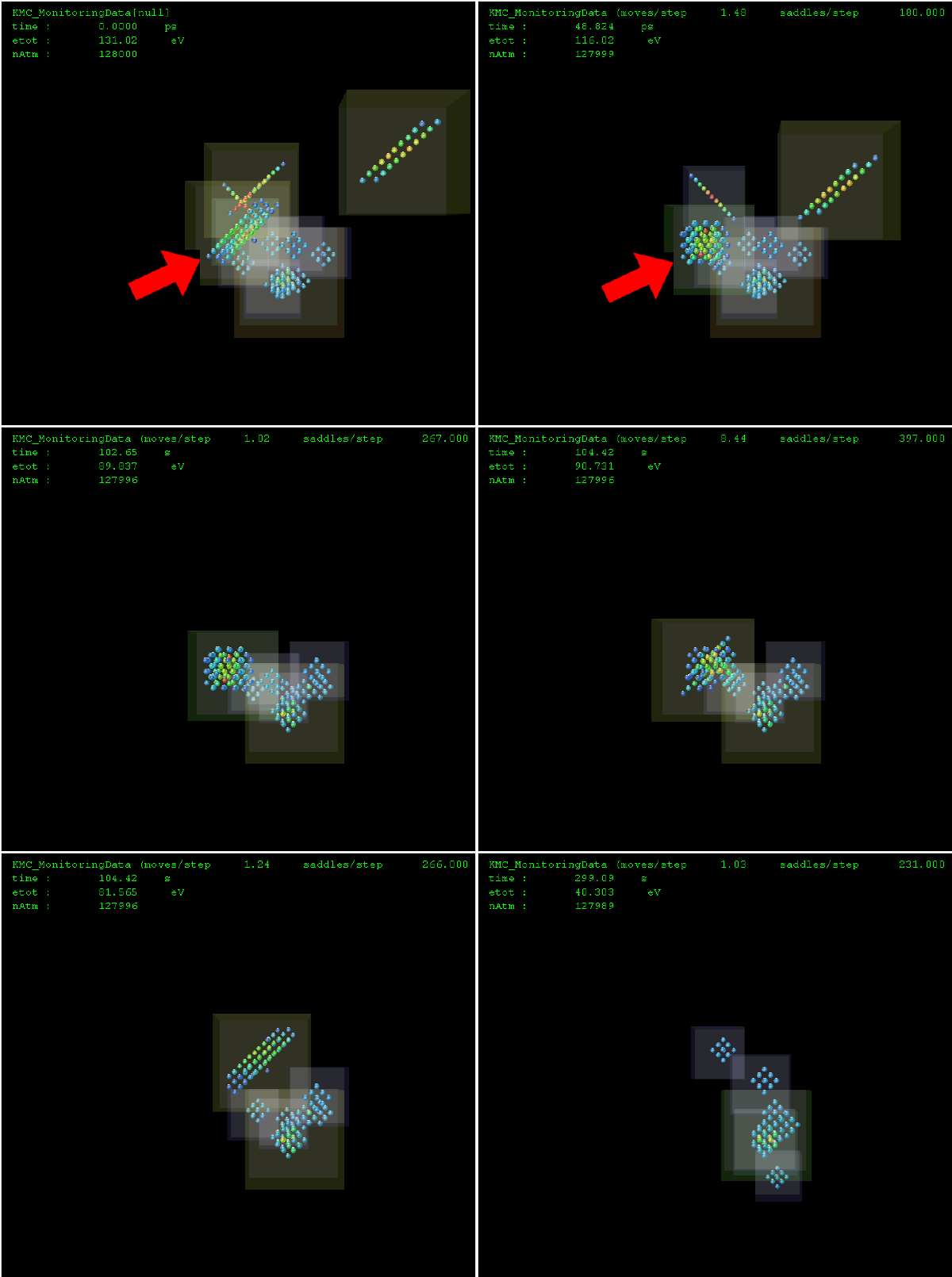}
               \caption{\label{fig:aokMC1}
               Snapshots from the evolution of a 20keV irradiation damage cascade in tungsten evolving at 600K.
               Reading left-to-right and top-to-bottom, the snapshots are taken at time $t=0, 50ps, 103s, 104s, 104s, 300s$.
               Starting from the top left image, the diinterstitial at the top left combines with the larger $6i\langle 111 \rangle$ interstitial cluster, indicated by a red arrow. These form a sessile 8-interstitial cluster, too small to identify a clear Burgers vector.
               The remaining monointerstitial and diinterstitial rapidly diffuse out of the simulation cell leaving only sessile objects.
               After 100s the 8-interstitial cluster absorbs a monovacancy, starting a rapid transformation into a mobile 7-intersitial $\langle 111 \rangle$ cluster.
               Brownian motion of the vacancies leads to accumulation into a single 9-vacancy void at about 1000s.
               Only high energy atoms are shown, coloured from blue ( $+0.1$eV ) through green to red ( $+1.5$eV ).
               This calculation took 58 h on a single processor.
                }
            \end{figure*}

            In figure \ref{fig:aokMC1} we show snapshots from a simulation showing the rapid formation and slow relaxation of a sessile interstitial cluster.
            We first see the collision $2i \langle 111 \rangle + 6i \langle 111 \rangle \rightarrow 8i \langle 100 \rangle$.
            This is a stable sessile defect, which would not then further evolve if simulated in isolation.
            In our case the sessile cluster is formed close to the centre of a dense cascade.
            The cluster is a complex object associated with a vacancy, too close to be treated separately. 
            After 100 seconds simulated time, the vacancy moves and is absorbed, in the reaction $8i \langle 100 \rangle + v \rightarrow 7i \mbox{(complex)}$.
            This then initiates a transformation into a glissile cluster $7i\mbox{(complex)} \rightarrow 7i \langle 111 \rangle$.
            In other simulations we have seen small sessile 3- and 4- interstitial clusters transform into a glissile cluster by absorption of an interstitial.
            A similar observation of vacancy-assisted detrapping was made previously using MD by Puigvi et al\cite{Puigvi_PM2007}.
            The difference here is quantitative rather than qualitative- our simulation was run at a temperature of 600K, where vacancies have a small mobility and so would never have been seen in MD.
        
            The last phase of annealing that we observe occurs after  1ms at 900K, or 1s at 600K.
            This phase is the onset of vacancy mobility.
            In the simulation shown in figure \ref{fig:aokMC2} the vacancies start well separated, and after one hour simulated time only a trivacancy and a quad-vacancy remain.

        \subsection{Discussion}
		\label{intClusters}

             \begin{figure*}
                \begin{minipage}{0.9 \textwidth}
                    \centering
                    \includegraphics[width=.9\linewidth]{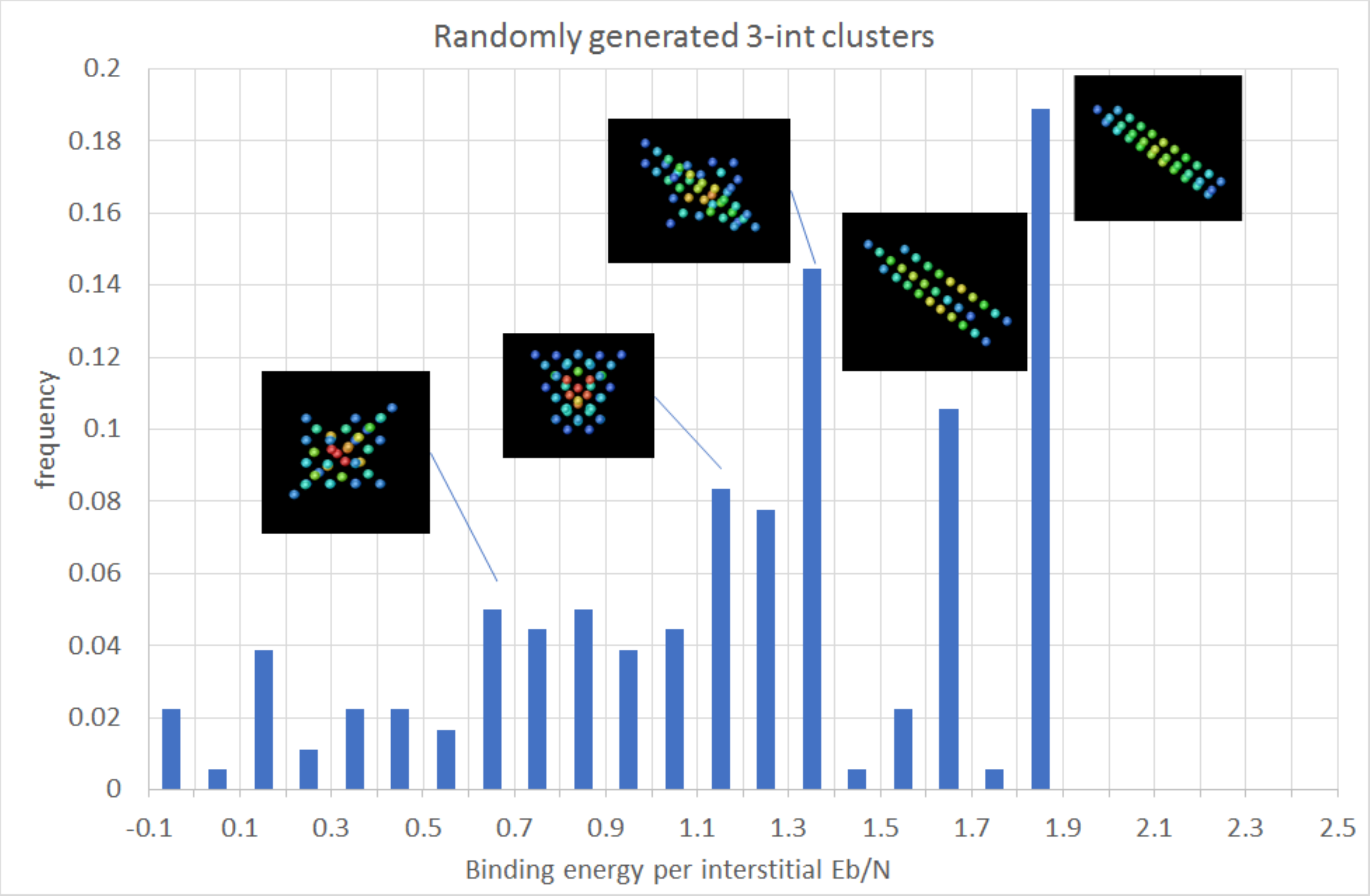}\\
                    \includegraphics[width=.9\linewidth]{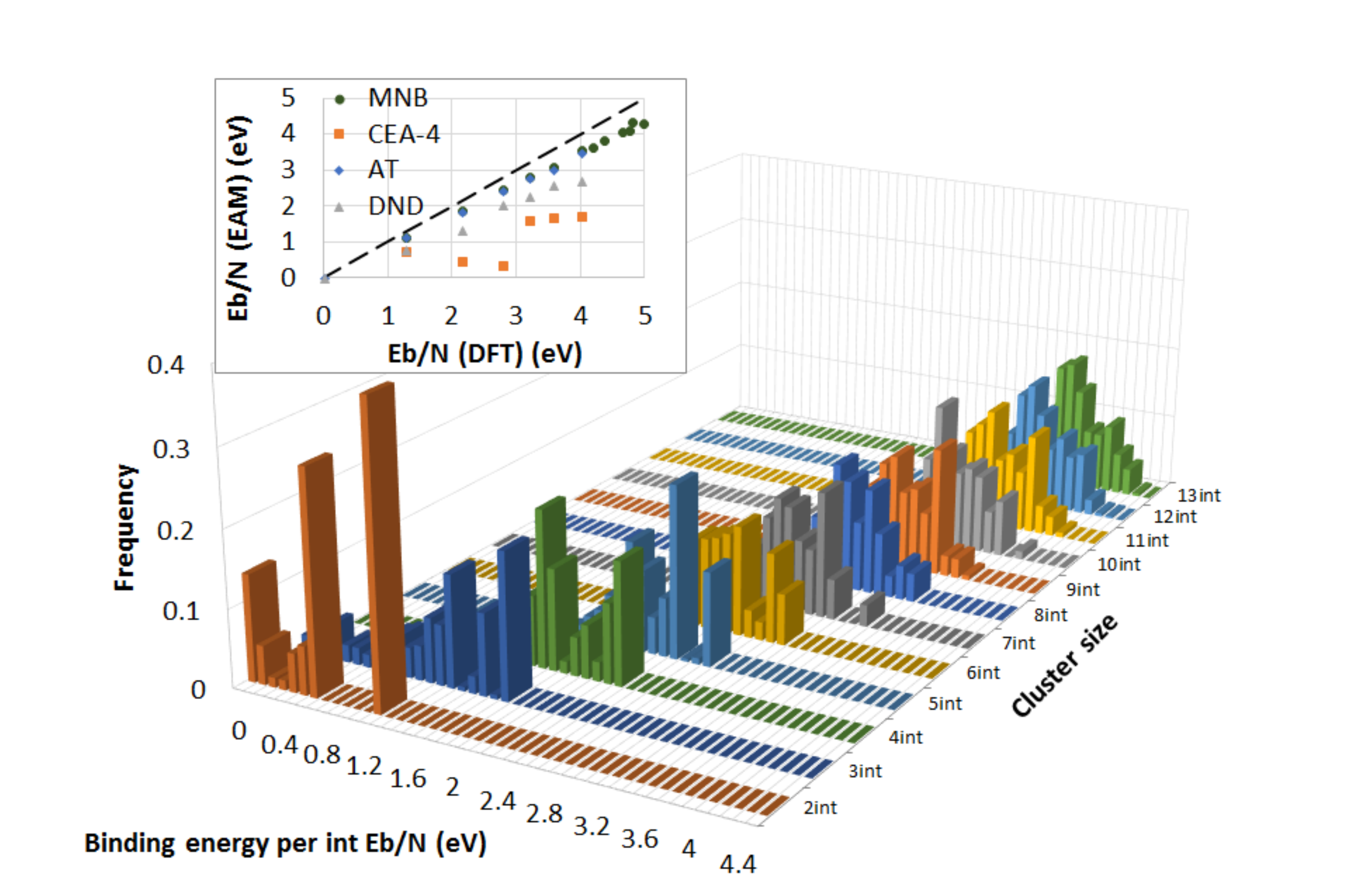}
                \end{minipage}%
               \caption{\label{fig:intClusters}
                If interstitials are randomly placed in close proximity and relaxed, a wide range of different structures are found.
                The top figure shows a histogram of binding energies found for three-interstitial clusters, with some of the structures drawn.
                The bottom figure shows similar histograms for cluster sizes 2-13 interstitials.
                Inset: a comparison of the binding energy for the minimum energy structures computed with DFT\cite{Alexander_PRB2016} and different empirical potentials.
                The lowest energy structures found for each cluster size are mobile $\langle 111 \rangle$-type defect clusters, but the majority are sessile and difficult to simply categorize.
                We conclude that if defects produced in cascades collide, they are likely to form a sessile metastable configuration before transforming to a mobile configuration.
                }
            \end{figure*}
            
The most physically interesting phase of the annealing is in the nanosecond to microsecond timescale, where complex sessile interstitial clusters are formed and subsequently relax, as these processes would not be observed in conventional object kinetic Monte Carlo with idealised object geometries.
To investigate the occurrence of sessile defect clusters, we have systematically explored the energy landscape of interstitial clusters in tungsten.
We placed $N$ interstitials into a $2 \times 2 \times 2$ unit cell box, randomly placed at the fine-mesh quarter lattice positions, then embedded this into a $16 \times 16 \times 16$ unit cell box and relaxed at constant volume.
We took $10 N^2$ initial configurations.
The binding energy is defined as the difference in formation energy between the lowest energy mono-interstitial and the $N$-interstitial cluster:
	\begin{equation}
    	E_b	\equiv N E^f_1 - E^f_N.
    \end{equation}
    
Figure \ref{fig:intClusters} shows the result as a histogram over binding energies. 
The height of the peaks is proportional to the number of times the bin is hit, without accounting for the same structure being found multiple times. 
We find that the lowest energy structures for each cluster size correspond to all interstitials close together and orientated along $\langle 111\rangle$, but there are also a large number of higher energy clusters, many of which are difficult to categorize.
This same conclusion about distinct low energy clusters and a semi-continuum band of metastable states was previously drawn by Marinica \textit{et al} for the Fe-interstitial landscape \cite{Marinica_PRB2011}.
A full study of these interstitial defect clusters can be found in ref \cite{Mason_2019}.
We conclude that the MNB empirical potential \cite{Mason_JPCM2017} gives good formation energies and relaxation volumes for these clusters, compared to DFT calculations.

The complex energy landscape of uncategorisable interstitial cluster configurations is significant for the microstructural evolution found in this study. 
This result suggests that if interstitial clusters generated in a cascade collide they can first form a single high-energy interstitial cluster, before subsequently relaxing further, ultimately to a low energy mobile defect cluster.

    \section{Conclusions}

        We have developed a new okMC code which takes atomistic configurations as the elementary objects, rather than using simple idealised defects.
        The code operates by searching for, and storing atomistic transitions, based on vacancy-atom exchange, and correlated atom rotation and string pulling moves.
        These moves are stored, and recalled, so that an object once discovered needs never be recomputed.
        Typically a large number of configurations of each defect cluster will be computed, but when all relevant configurations are found no further expensive atomic searches or relaxations are required and the code speed is comparable to a regular okMC code, albeit with a higher memory footprint.
        If defects collide, or transform into a previously unexplored configuration, the object can not be recalled from the database and so more atomistic transitions are added.

        Elastic interactions are treated differently at four different length-scales.
        Within an object, the atoms are kept relaxed using interatomic potentials, allowing for arbitrarily complex atomic configurations.
        When a local transition is considered, this is done by clamping atoms on the boundary of the transition active volume.
        The energy impact of this clamping is removed by computing the harmonic elastic relaxation energy, computed using the Hessian.
        The self-energy due to interactions with periodic replicas is subtracted using the dipole tensor (computed with interatomic potentials) and the isotropic elastic Greens function, in the manner of Varvenne et al\cite{Varvenne_PRB2013}.
        Finally the interaction between objects is treated using dipole tensors and isotropic elasticity.
        The long-range elastic energy difference between the before- and after- states is added to each transition.

        We have found that the first nanosecond simulated time is by far the most computationally expensive part of the simulation.
        In future work it would be advantageous therefore to exploit MD to run the first nanosecond of simulation, rather than stopping the simulation shortly after the PKA was set in motion.
        Our cascade annealing results were produced using a serial code, taking tens of hours to complete each run.
        While it may be possible to parallelize the code in the future, this is not a trivial task due to the difficulty in load-balancing: each kMC event requires atomic relaxations with very different times required for convergence.
        As our computational burden is light, we can of course benefit from the trivial parallelization offered by running multiple independent simulations for generating statistics.
        It is also important to note that we have deliberately simulated cascade annealing in the dilute (low-fluence) limit, exploiting the loss of mobile defects to distant sinks to accelerate the dynamics. 
        In the dense limit we would expect cascade overlaps, which would be better modelled with MD rather than kMC alone.
        
        Our results demonstrate that there may exist interesting modes of cascade relaxation, where glissile defects collide to form a sessile defect.
        This sessile defect may then spontaneously transform to a glissile form, or may have such a transformation initiated by absorption of a mobile defect.
        At present we do not have sufficient statistical information to be able to draw strong conclusions about the relative rate of these relaxation modes, but this does indicate that it may be necessary to take into account the spatial correlation of multiple defects generated in a small cascade.

    \section{Acknowledgements}

        The authors would like to thank Max Boleininger, CCFE and Fredric Granberg, University of Helsinki, for helpful discussions.
        This work has been carried out within the framework of the EUROfusion Consortium and has received funding from the Euratom research and training programme 2014-2018 and 2019-2020 under grant agreement No 633053, and from the RCUK Energy Programme [grant number EP/P012450/1].
        The views and opinions expressed herein do not necessarily reflect those of the European Commission.
        AES acknowledges support from the Academy of Finland through project No. 311472.
        To obtain further information on the data and models underlying this paper please contact PublicationsManager@ccfe.ac.uk.

    \bibliographystyle{unsrt}
    \bibliography{aokmc}

\end{document}